\documentclass[]{article}

\begin{document}
\begin{center}
\Large SIR JAMES JEANS AND THE STABILITY OF GASEOUS STARS

\vspace{1.0cm}

\normalsize
{\it by Alan B. Whiting \\
University of Birmingham}
\end{center}

\vspace{1.0cm}

\large
In 1925 Sir James Jeans calculated that a star made up of an ideal
gas, generating energy as a moderately positive function of 
temperature and density, could not exist.  Such stars would be
unstable to radial oscillations of increasing size.  It appears that
the flaw in his calculation has never been clearly
explained, especially the physical basis for it. 
I conclude it lies in an almost offhand assumption made 
about the form of the temperature perturbation.  The episode 
provides a number of lessons about complicated calculations and
their interpretation.
\normalsize

\vspace{1.0cm}

\noindent {\it What is a Star?  The view in 1925}

In the first decades of the twentieth century one of the areas of great
astrophysical interest and activity was stellar structure.  It was fairly
clear that the Kelvin-Helmholtz theory of gravitational contraction
(which lives on in our terminology of `early-type' and `late-type' stars)
was inadequate, though its replacement was not immediately available.
The theory of relativity promised in principle a sufficiently
long-lasting power source through the conversion of mass to energy, but
details remained obscure.  Using the nineteenth-century science of
thermodynamics and the twentieth-century theories of radiation,
models of gaseous stars were
constructed and elaborated, eventually forming the basis of the
science of stellar structure as we know it today.

But of course it was not clear at the time that such an approach would
necessarily work.  In fact Sir James Jeans, a major figure in astrophysics,
put forward a theory involving liquid stars powered by a form of nuclear
fission.  A complete treatment of his theory and its fate is, however, beyond 
the scope of a short paper like this one.  Here I intend to treat just one
aspect of Jeans' work: his calculation that gaseous stars were unstable.

With the benefit of hindsight we can say that Jeans' calculation must have
been in error.  There are several possible explanations, from
simple mistakes in algebra 
to new physics that was simply unknown in 1925.
My initial aim in this study is to determine where Jeans' error lies, and 
especially whether it could have been detected using methods and knowledge
available to him.  Subsequently I will examine some of the implications
of the episode for the practice of mathematical model-building and for
the process of science.

It is possible to make many more connexions between Jeans' ideas and 
subsequent work, but for limitations of space I will maintain the focus
on his calculation and things having a direct bearing on it.

\vspace{0.5cm}
\noindent{\it Jeans' Analysis}

Jeans$^1$ presented his results in a {\it Monthly Notices} paper in 1925,
but I will use as a reference the revised version$^2$ published
in book form a few years later.  The relevant parts are sections 105-111,
found on pages 117 through 125.  I will outline them here, referring the
reader to the original for the details.

Jeans begins by writing the equation of motion for a shell at radius
$r$ within a star, supported by pressure and held together by gravity:
\begin{equation}
\frac{d^2 r}{dt^2} = -\frac{1}{\rho} \frac{d}{dr} 
\left(p_G + \frac{1}{3} a T^4 \right) - \frac{\gamma M_r}{r^2}
\label{dyn1}
\end{equation}
where $\rho$ is the mass density, $p_G$ the gas pressure, $a$ the radiation
constant, $T$ the temperature, $\gamma$ Newton's constant and $M_r$
the mass within the radius $r$.
Next Jeans accounts for energy raising the temperature of the material,
doing pressure-volume work, being generated by some
process as yet unspecified, and flowing out and in:
\begin{equation}
\rho C_v \frac{dT}{dt} - 
\left(p_G + \frac{4}{3} a T^4 \right)\frac{1}{\rho} \frac{d \rho}{d t} =
\rho G - \frac{1}{r^2} \frac{d}{dr} \left( r^2 H \right)
\label{heat1}
\end{equation}
where $C_v$ is the heat capacity of the material, $G$ the rate of 
generation of energy per unit mass, and $H$ the radiant energy flux.

Next the model is made more specific by assuming something like an
ideal gas
\begin{equation}
p_G \propto T \rho^{1+s}
\end{equation}
where $s$ may be used to parametrize departures from ideal gas
behaviour; and an opacity similar to that of Kramers' expression
\begin{equation}
k = \frac{c \mu \rho}{T^{3+n}}
\end{equation}
with $k$ the coefficient of opacity, $\mu$ the molecular weight of
the material and $c$ the speed of light.  If $n = 1/2$ a Kramers
opacity is recovered.  Assuming an Eddington grey atmosphere, this
expression allows us to write the equilibrium radiation flux $H$
in a useful form.

Jeans assumed an energy generation law in which
\begin{equation}
G \propto \rho^\alpha T^\beta
\end{equation}
and, for convenience in notation, introduced $\lambda$ as the
ratio of gas pressure to radiation pressure.

To this point Jeans has built a star almost as one would do today.  
It is of course much
simpler than the models one builds nowadays on a computer, but
there should be nothing inherently unstable about these simplifications.

The next step is the stability analysis.  The general problem of
stability for a system as complicated as this is difficult and
complicated to treat analytically.
Through various relations and assumptions
Jeans reduced the problem to that of the perturbation in size, 
$\delta r$; and assumed
a particular form of perturbation, a proportional change in radius
(so that $\delta r/r_0$ is constant with radius).
This still left too many terms in $\delta T/T_0$, so Jeans
assumed also that it was constant with radius; that is, there was
a proportional heating or cooling everywhere.  I think this is
a very important assumption, though it does not seem to have
excited any comment before now. 

With these assumptions and simplifications made, Jeans obtained
the master equation of stability:
\begin{eqnarray}
\frac{d^3}{d t^3} \delta r & + &\frac{(7+n-\beta)G_0}{C_v T_0}
\frac{d^2}{d t^2} \delta r \nonumber \\
& + &\frac{\gamma M_r}{r_0^3} \left[
\frac{\lambda +4}{\lambda+1} \left(\frac{3p_G+4aT_0^4}{\rho C_vT_0}
-1 \right) + \frac{3 s \lambda}{\lambda + 1} \right] \frac{d}{dt}
\delta r  \nonumber \\
& + & \frac{\gamma M_r}{r_0^3} \frac{G_0}{C_v T_0} \left[
\frac{\lambda +4}{\lambda+1} \left(3 \alpha + \beta - n \right)
+ \frac{3 s \lambda}{\lambda + 1} \left( 7 + n - \beta \right)
\right] \delta r = 0.
\label{stab1}
\end{eqnarray}
If one cares to reproduce the algebra, one finds also that Jeans
has assumed that $\lambda$ does not vary with radius, nor $C_v$.
In any realistic star they will; but not so much as to change the
qualitative stability.

Equation (\ref{stab1}) is a linear equation with constant coefficients
(constant with respect to time; in general they may vary from shell to
shell)
so the solutions will be of the form $e^{wt}$.  There are three possible
values for $w$, solutions of the equation
\begin{equation}
w^3 + B w^2 + C w + D = 0
\label{aux}
\end{equation}
where $B$ is the coefficient of $d^2 \delta r/d t^2$ in Equation
(\ref{stab1}), and similarly.  Following Jeans
we note that any positive $w$, or complex $w$ with a positive real
part, denotes instability.  Avoiding a real, positive $w$ gives the
condition that $D$ 
be positive.  If we assume a perfect gas, so $s=0$, this means
\begin{equation}
3 \alpha + \beta - n > 0.
\label{stab2}
\end{equation}
So if the coefficients of energy generation are too small, the star
will monotonically shrink or expand.  Physically, this means that if
compressing the star (say) does not generate enough additional heat
to cause the pressure to rise and bounce back, the star keeps on
shrinking.  This is reasonable, and so far we have no argument with
Jeans' calculations.

Avoiding a complex $w$ with a positive real part is a bit more 
involved, but if $D$ is positive the condition reduces to
\begin{equation}
BC > D
\label{stab3}
\end{equation}
which means (for $s=0$)
\begin{equation}
\left(7+n-\beta \right) \left( \frac{3p_G+4aT_0^4}{\rho C_v T_0}
-1 \right) > 3 \alpha + \beta - n.
\label{stab4}
\end{equation}
(The expression Jeans gives is a bit different,
but in the last step of his derivation
he appears to have made a substitution that is
true only for critical stability, that is, only when the inequality
is an equation.  For purposes of finding the critically stable
exponents no trouble should have resulted, however.)

Taking Equation (\ref{stab4}) together with Equation (\ref{stab2}),
we find that the fraction in the brackets (something like the ratio
of pressure-energy to thermal energy) must be greater than unity.
For a star in which only gas pressure is important it has the
value 3/2, while if radiation pressure dominates it approaches 1.
At unity, the two stability criteria together require
\begin{equation}
3 \alpha + \beta - n = 0
\end{equation}
an extremely restrictive condition; if a two-particle fusion reaction is
postulated as an energy source so that $\alpha = 1$, along with a Kramers
opacity, it must go {\em slower}
as temperature increases, and at an exact rate.  
The situation with a ratio of 3/2 is better,
but still a two-particle reaction cannot depend even linearly on
the temperature.  Jeans' stability criteria exclude essentially all
nuclear fusion reactions as a possible source of stellar energy.

\vspace{0.5cm}
\noindent{\em Oscillations and the Thermal Instability}

What is the {\em physical} cause of Jeans' unstable stars?  
He interpreted the oscillations of increasing amplitude
as being due to the increased heat-energy liberated by the reaction 
(whatever it is) during the dense phase of the oscillation requiring an
expansion of a larger amplitude during the following phase.  This is a
plausible interpretation, but worth looking at in more detail.

The perturbed energy equation (from Equation \ref{heat1}) is
\begin{eqnarray}
\rho C_v T \frac{d}{dt} \left( \frac{\delta T}{T} \right)& -&
\rho G \left( \beta -7-n \right) \frac{\delta T}{T} = \nonumber \\
&-& 3 \left( C_v T \rho^{1+s} + \frac{4}{3} a T^4 \right)
\frac{d}{dt} \left( \frac{\delta r}{r} \right) - \rho G
\left( 3 \alpha + 7 \right) \frac{\delta r}{r}
\label{heat2}
\end{eqnarray}
where the subscript noughts, meaning equilibrium values, have
been suppressed for ease of notation.
Now suppose we impose a sinusoidal oscillation in $\delta r /r$.
The equation may now be thought of as a linear, first-order
differential equation in the (proportional) temperature 
perturbation with a
sinusoidal forcing function.  The solution will be the sum of
sinusoidal terms and the homogeneous solution, which is
\begin{equation}
\frac{\delta T}{T} \propto \exp \left( \frac{G}{C_v T} 
(\beta -7-n) t \right).
\label{temp1}
\end{equation}
In order for the temperature to stay bounded, $\beta \leq 7+n$.
This is not as strict a requirement as Jeans found, but then this
is a {\em different} calculation: we have found the temperature
response to a forced radial motion of the star.
It still puts an upper limit on the temperature sensitivity of the reaction,
and in particular excludes the exponential dependence one expects for
nuclear fusion reactions.  Jeans' interpretation
is thus shown to be correct.  In addition, having a physical
source for Jeans' instability allows us to investigate more
complicated situations.

\vspace{0.5cm}
\noindent{\em Previous Identifications of the Flaw}

For a calculation that appeared to show gaseous, fusion-powered stars
impossible, Jeans' work has left very little lasting trace.
Chandrasekhar's$^3$ 1939 work on stellar structure, very 
extensive in its listing of the literature, does not even mention it
directly.  He does refer to an unspecified `belief' in instability
resulting from a certain kind of energy-producing
reaction (pp. 457, 468), but sees no
convincing reason for it.
Milne's$^4$ biography of Jeans describes the calculation in
some detail, but as far as criticism notes only (p. 138) 
that in the calculation
he parametrises departures from the ideal gas laws in two
incompatible ways.  This does show that Jeans was not terribly
interested in specifically how stars might depart from being
ideal gases, but then Jeans 
says just that himself at the beginning of his calculation.  
At any rate, it has no effect on the
stability analysis for ideal-gas stars. 
(It is possible that Milne thought Jeans' work
might be essentially correct; in his description he shows
scepticism about the gaseous model.)
The much more recent work on pulsating stars by Cox$^5$ does mention
Jeans' calculation (pp. 166 and 172).  It is noted as being 
equivalent to a one-zone model of a star (an important point, and
one discussed below), but no analysis of its flaw is given.

What appears to be the accepted answer for the flaw in Jeans'
calculation appears in a pair of papers by Cowling$^{6,7}$
(the first of which is cited by Chandrasekhar).
He pointed out that a proportional change in radius, with
$\delta r/r$ constant in space, probably was not a normal
mode for a star.  Therefore (putting words in his mouth) giving it
such a kick would excite a number of normal modes of different
frequencies, and the perturbation would not remain a constant
proportion.  Jeans' perturbation calculation was then not strictly a
{\em stability} determination, in which the star is left to itself
after being kicked, but shows a response to a specific kind of
forcing.

Cowling had found a flaw in Jeans' work that made its result
questionable.  But this is not the same thing as showing that
the flaw in fact led to the erroneous conclusion.  Indeed, it
is not immediately clear how a combination of normal motions, each of them
individually stable, should be unstable taken together.
(More recently, Papaloizou$^{8,9}$ has shown that 
at least for some models of very
high-mass stars, the opposite can happen: mode mixing can
stabilise otherwise unstable modes.)

Cowling did go on to impose a proportional radial oscillation on
his analysis, and obtained essentially Jeans' result.  But because
he was dealing with oscillations assumed to be adiabatic or nearly
so, the temperature perturbation was automatically of the same form
as the radial perturbation;
so Jeans' separate assumption about it was implicitly included.
It is this assumption concerning the temperature perturbation that
I think is the real problem with Jeans' analysis.  The next step is
to test this idea.

\vspace{0.5cm}
\noindent{\em Non-Proportional Temperature Perturbations}

The straightforward way to check the effect of this assumption is
to relax it.  One perturbs Equations (\ref{dyn1}) and (\ref{heat1})
again, this time allowing terms with various derivatives
of $(\delta T/T)$ with radius, then collecting everything into a new
and more accurate version of Equation (\ref{stab1}).  In principle
it is possible.  But instead of requiring one additional equation,
the time derivative of Equation (\ref{dyn1}), and a bit of algebra,
one needs nine additional equations and a truly horrible lot of
algebra.  It is not a practical pen-and-paper exercise and I do not
believe Jeans would have seriously considered it.

Even writing down Equation (\ref{heat2}) as it appears with
the relaxation of the constraint on the temperature perturbation
is not particularly enlightening.  But by making the further assumption
that the temperature perturbation is separable, so
that
\begin{equation}
\frac{\delta T}{T} = R(r) X (t).
\label{separate}
\end{equation}
the expanded form of Equation (\ref{heat2}) can be rearranged into
\begin{eqnarray}
&&\rho C_v T RX' + \nonumber \\
&&  \left[  -\rho G \left( \beta -7-n \right) R  - \left(
(7+n) H +
 \frac{1}{r^2}\frac{d}{dr} \left( \frac{r^2TH}{dT/dr} \right) 
 \right) R'
 - \frac{HT}{dT/dr} R'' \right] X \nonumber \\
&&= - 3 \left( C_v T \rho^{1+s} + \frac{4}{3} a T^4 \right)
\frac{d}{dt} \left( \frac{\delta r}{r} \right) - \rho G
\left( 3 \alpha + 7 \right) \frac{\delta r}{r}
\label{heat4}
\end{eqnarray}
where primes denote derivatives with respect to the independent
variable.
If we impose a sinusoidal variation in the radial perturbation as
before and consider this an equation for $X$, we again have a
first-order differential equation with a forcing term.  This
time the coefficients depend on the variation of the temperature
perturbation with radius, which we do not know.  But some qualitative
reasoning about the relative sizes of the $R$, $R'$ and $R''$
coefficients shows that the radial change in temperature perturbation
need only be a small fraction of the perturbation itself to dominate
the stability calculation.

Physically, Jeans' assumption of a strictly proportional temperature
perturbation has the effect of pumping heat energy around the star in
an unphysical way.  It is not a large effect, taking many oscillations
to grow significantly (so that the adiabatic assumption of Cowling
and many others is a very good one for working out small
pulsations), but enough to blow up the star eventually.

On a more abstract level, it appears that the problem with Jeans'
analysis lies in ignoring the internal structure of the star, in
requiring it to be in some sense a single unit.  We may explore
this idea with a toy star.

\vspace{0.5cm}
\noindent{\em The Toy Star}

Not as a serious mathematical model, but as a way of investigating
some of the effects we expect to operate in a real star, I present
here a very simple model.  (This is similar in motivation and
in general to that in Kippenhahn and Weigert$^8$, pp. 13-15,
235-8 and 407-8, though the details and the application are different.)
Suppose we enclose a spherical quantity of ideal gas, of mass $m$ and
density $\rho$, with some mixture of radiation
inside a hollow shell of radius $r$ and mass $M$.  The gas mass
$m$ is much smaller than $M$; the shell is supported by the 
pressure of gas and radiation, and is held in by gravity.
Everything is at one temperature $T$  and the surface radiates
heat as a blackbody.  Inside the shell there is an energy-generating
reaction that goes as $\rho^\alpha T^\beta$.  The equations of
motion and energy balance are thus
\begin{eqnarray}
M\frac{d^2 r}{dt^2} & = & - \frac{\gamma M^2}{r^2} + \left( p_G +
\frac{1}{3} a T^4 \right) 4 \pi r^2 \nonumber \\
m G & = & m C_v \frac{dT}{dt} +
\left(p_G + \frac{4}{3} a T^4 \right) 4 \pi r^2 \frac{dr}{dt}
+ 4 \pi r^2 \sigma T^4.
\label{piston}
\end{eqnarray}
Perturbing these, the heat balance equation becomes
\begin{equation}
U \frac{d}{dt} \left( \frac{\delta T}{T} \right) =
\left( \beta - 4 \right) F_0 \frac{\delta T}{T} -
\left( \lambda + 4 \right) \frac{4 r}{3c} F_0 
\frac{d}{dt} \left( \frac{\delta r}{r} \right)
- \left( 3 \alpha + 2 \right) F_0 \frac{\delta r}{r}
\label{heat6}
\end{equation}
where $U= \rho C_v T$ is the (unperturbed) thermal heat content and
$F_0 = 4 \pi r^2 \sigma T^4$ is the (unperturbed) energy radiated away
from the surface.  The master stability equation is derived as before.
The first condition for stability, avoiding a monotonic
expansion or collapse, implies
\begin{equation}
3 \alpha + \beta - 2 > 0
\label{stab8}
\end{equation}
a lower limit on the energy-generating reaction.  The second
criterion reduces to
\begin{equation}
\left(4-\beta \right) \left( \lambda + 4 \right) \frac{4rF_0}{3cU}
> 3 \alpha + 2.
\label{stab9}
\end{equation}

The fraction can be interpreted as $F_0/U$, the reciprocal of
the time required to radiate away internal heat energy without
replacement; times 4/3 the time required for light to go from the
centre of the sphere to the surface.  For even a toy star the
ratio of crossing time to radiative timescale
should be very, very small, giving us almost
independently of the value of $\beta$,
\begin{equation}
\alpha  < - \frac{2}{3} 
\label{stab10}
\end{equation}
that is, that the energy-generating reaction is required to go
{\em more slowly} with increasing density.

If we look again at Equation (\ref{heat6}) and impose a sinusoidal
radial oscillation, we find that the temperature perturbation
response is given by a sinusoidal term plus
\begin{equation}
\frac{\delta T}{T} \propto e^{\left( \beta-4 \right) \frac{F_0}{U} t}
\label{driven2}
\end{equation}
that is, if the energy generation reaction produces energy faster
than it can be radiated away, temperature increases
exponentially on a radiative timescale.

Taken together, these results show the toy star behaving in
a similar way to Jeans' gaseous stars, and for similar reasons.
The energy generation rate must be sensitive to temperature and
pressure, but in an extremely restrictive and unphysical way.
Too little sensitivity and the star implodes or explodes monotonically;
too much and growing oscillations, due to excess heat energy,
tear the star apart.

\vspace{0.5cm}
\noindent{\em The Stability of Gaseous, Fusion-Powered Stars}

It is important to keep in mind just what all these calculations have,
and have not, proven.  Cowling called attention to the fact that
(in my words)
Jeans' analysis was not strictly a stability calculation, since the
required form of perturbation was not a free one.  Since he recovered
Jeans' result by imposing Jeans' restrictive form of the radial
perturbation, he concluded that this was the problem.  It is true enough
for most purposes, since pulsations are very nearly adiabatic and for
those the perturbations are proportional.
By focussing on the
temperature perturbation I have shown that Jeans' physical explanation
for his instability is correct, that is, that each
oscillation (of the restricted type) produces more heat energy than
steady motion dissipates.  I have then shown that 
even a small departure from a strictly proportional temperature
perturbation could plausibly stabilise a forced 
oscillation, so the physical mechanism that
destabilises a Jeans star does not operate.  Thus a full stability
analysis using his methods (possible in principle, though not likely
in practice) would show that gaseous, fusion-powered
stars are stable. 
Since Jeans' assumptions of the form of temperature
and radius perturbations in some sense ignore the structure of the
star, I constructed a structureless toy star, and found it to be
subject to the same kind of instability that Jeans found, thereby
lending support to that interpretation.

In all this I have {\em not} proven the stability of gaseous,
fusion-powered stars.  Indeed it seems almost certain that such proof
is beyond the specific techniques used by Jeans.
Cowling$^{6,7}$ took a more in-depth approach,
depending more on the details of gaseous stars (some of which had
not been determined when Jeans wrote).  In that sense he was less
general; but in allowing any shape to the perturbations he was
more general in the more important way.  In the end, his work was
accepted as the true answer.

It appears that the problem was inherently more
complicated than Jeans allowed for, not allowing of reduction to a
single ordinary differential equation.  It was, however, possible
for him
to see that relaxing the assumption concerning the temperature
perturbation {\em could} have changed the conclusion, and thus that
the calculation was actually inconclusive.  

\vspace{0.5cm}
\noindent{\em Lessons for Mathematical Modelling and the Progess
of Science}

Out of all this algebra has come one potentially useful insight into
the structure of stars: a star {\em must} change its structure to be
stable.  Simple proportional 
expansion and cooling, or contraction and heating, is
not quite enough.  Our concern, however, is much more with the
implications of the episode for the practice of science.

Jeans' mistaken calculation had very little direct effect on the progress of
stellar modelling.  Its main importance for us here lies, first, in its
implications for the continuing practice of constructing mathematical
models of astrophysical systems.
The actual flaw was subtle in its introduction and effect and went
undetected for a very long time.  The lesson that mistakes happen and
that they may not be found very soon is not a new one, but well worth
underlining. 
Also worth emphasising is the fact that every mathematical assumption has
some physical implication and that the true influence of a simplifying
assumption is not known until it is relaxed.  Models have grown no simpler
since Jeans' day, but simplifying assumptions with all their effects
must still be made.  

This episode is important, secondly, for its implications about the
progress of science, {\em exactly} because it had little effect.  
A calculation that appeared to disprove a popular
picture (it hardly yet amounted to a theory) made little impact, in
spite of the fact that its flaw was not discovered for a decade.
Indeed, one may say the flaw was never actually found, but
rather a different (more complicated and more correct) calculation was
performed that replaced Jeans' effort.
Note that Cowling
was investigating in great detail the stability of {\em
gaseous} stars, not the liquid versions postulated by Jeans. 
This episode, then, illustrates the process of accepting a new theory
as noted in Kuhn's$^{11}$ original work (notably
chapter XII): the new theory
need not answer all questions at once, and indeed might not even do as well
as another theory in specific places.  A theory is almost
necessarily a vague and imperfect thing at the beginning.  It is only
after a long period of refinement that its true power is shown.

The author is grateful to an anonymous referee for many helpful suggestions,
and especially for bringing to his attention several relevant
papers and passages in the literature.

\vspace{1.0cm}
\begin{center}
{\it References}
\end{center}

\noindent (1) J. H. Jeans, {\em MNRAS}, {\bf 85}, 914, 1925\\
\noindent (2) J. H. Jeans, {\em Astronomy and Cosmogony (2nd Edn.)}
(Dover, New York), 1961; unaltered reprint of the 1929 version \\
\noindent (3) S. Chandrasekhar, {\em An Introduction to the Study of
Stellar Structure}, (Dover, New York), 1958; unaltered reprint of
the 1939 version \\
\noindent (4) E. A. Milne, {\em Sir James Jeans}, (Cambridge University
Press), 1952 \\
\noindent (5) J. P. Cox, {\em Theory of Stellar Pulsation},
(Princeton University Press), 1980
\noindent (6) T. G. Cowling, {\em MNRAS}, {\bf 94}, 768, 1934 \\
\noindent (7) T. G. Cowling, {\em MNRAS}, {\bf 96}, 42, 1935 \\
\noindent (8) J. C. B. Papaloizou, {\em MNRAS}, {\bf 162}, 143, 1973 \\
\noindent (9) J. C. B. Papaloizou, {\em MNRAS}, {\bf 162}, 169, 1973 \\
\noindent (10) R. Kippenhahn and A. Weigert, {\em Stellar Structure and
Evolution}, (Springer-Verlag, New York), 1991 \\
\noindent (11) T. S. Kuhn, {\em The Structure of Scientific Revolutions
(2nd Edn., enlarged)}, (University of Chicago Press, London, Chicago),
1970
\end{document}